# Analyzing Social Media Data to Understand Consumers' Information Needs on Dietary Supplements

Rubina F. Rizvi[a,b]*, Yefeng Wang[a] *, Thao Nguyen[c], Jake Vasilakes[a,b], Jiang Bian[d], Zhe He[d], Rui Zhang[a,b]

[a] Institute for Health Informatics, University of Minnesota, Minneapolis, MN, USA,
[b] Department of Pharmaceutical Care & Health Systems, University of Minnesota, Minneapolis, MN, USA
[c] Data Science, University of Minnesota, Minneapolis, MN, USA
[d] Department of Health Outcomes & Biomedical Informatics, University of Florida, Gainesville, FL, USA,
[e] School of Information, Florida State University, Tallahassee, FL, USA,

(*Equal contributor first authors)

**Abstract**

*Despite the high consumption of dietary supplements (DS), there are not many reliable, relevant, and comprehensive online resources that could satisfy information seekers. The purpose of this research study is to understand consumers' information needs on DS using topic modeling and to evaluate its accuracy in correctly identifying topics from social media. We retrieved 16,095 unique questions posted on Yahoo! Answers relating to 438 unique DS ingredients mentioned in sub-section, "Alternative medicine" under the section, "Health". We implemented an unsupervised topic modeling method, Correlation Explanation (CorEx) to unveil the various topics consumers are most interested in. We manually reviewed the keywords of all the 200 topics generated by CorEx and assigned them to 38 health-related categories, corresponding to 12 higher-level groups. We found high accuracy (90-100%) in identifying questions that correctly align with the selected topics. The results could be used to guide us to generate a more comprehensive and structured DS resource based on consumers' information needs.*

## Introduction

Dietary supplements (DS) usage has gained popularity in recent years with almost 52% of U.S. adults reporting the use of one or more supplement [1]. This high DS usage is especially common among adults aged ≥60 years, where 70% have reported using one or more DS [2]. In spite of this escalating trend in DS consumption across a wide range of consumers, there are not many online resources out there that consumers could refer to for DS information that is personalized, reliable, complete yet succinct, up-to-date and in a language that is easily comprehensible by a lay-person.

In recent years, the internet has emerged as an important source of health-related information providing an opportunity for people to search online for free health information. According to a Pew Research Center report, 80% of internet users have looked for health information online [3, 4]. This would be especially true in the case of DS as its use is primarily self-initiated rather than based on clinicians' recommendations [5]. Existing online DS health information resources in the U.S. can range from open access, publicly available databases, e.g., Food and Drug and Administration (FDA) [6]; Office of Dietary supplements (ODS) [7]; Dietary Supplement Label Database (DSLD) [8], to commercial databases that often require a paid subscription, e.g., Natural Medicines (NM) [9]. When it comes to personalized queries from consumers, they are often consolidated under online resources as frequently asked questions. However, the information dissipated from such resources is often very basic, non-specific, and not very helpful.

The rapid growth of digital data in today's world, especially in the healthcare domain, offers great opportunities for its secondary use in clinical research. Topic modeling [10] has been an area of great interest and to date, several studies have been conducted to make use of electronic data and utilize this novel methodology. The reason for its growing popularity is its ability to reveal the latent structure and groupings of the underlying corpus without any prerequisite knowledge. Some of the applications of topic modeling in healthcare research include: analyzing clinical notes from Electronic Health Record (EHR) data; discovering and understanding health care trajectories [11]; identifying medication prescribing patterns [12]; mining adverse events of DS from product labels [13]; and discovering health topics in social media [14, 15] among various others.

There are various social Questions and Answers (Q&A) sites and online forums within health communities, e.g., Yahoo! Answers, allowing one to seek information through posting questions and receiving answers from others users (e.g., consumers, health professionals) [16]. Previously, we have used Yahoo! Answers data in several studies e.g., to investigate the terminology and language gap between health consumers and health professionals [17]; to mine consumer friendly medical terms to enrich consumer health vocabulary [16]; and to understand the information needs for diabetes patients about their laboratory results [18].

The purpose of this research study is to understand the information needs of DS consumers by analyzing questions coming directly from consumers and in their own language. The goal is achieved by using Correlation Explanation (CorEx) - a topic modeling algorithm on the title and body of each question under the Q&A section of the Yahoo! Answers database in order to unveil the "topics" around DS information needs. We generated a list of coherent topics that more accurately represent the areas of DS-related information and associated DS ingredients that consumers are most interested in. We will also evaluate the accuracy of the CorEx method in correctly identifying the topics from social media. In the future, the knowledge gained from this study could be used as a guide for developing more meaningful DS resources for consumers that are better aligned with their information needs.

## Methods

Figure 1 illustrates the overview of the methods. We extracted and pre-processed questions retrieved from the Yahoo! Answers database, focusing on questions around DS. We performed topic modeling using CoRex in order to understand DS-related topics and categories that consumers are most interested in. We then evaluated the accuracy of the topic modeling methodology by manually reviewing a subset of top ranked questions. We further investigated the actual DS ingredients associated with all the questions under each topic.

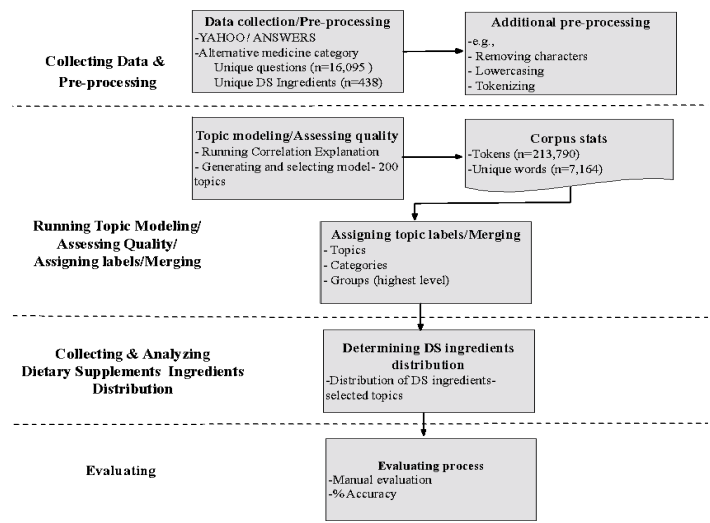

*Figure 1-Process overview*

### Collecting and Pre-processing Data

We collected in total 2,820,179 Yahoo! Answer questions and their corresponding answers posted under 21 sub-categories belonging to the main category "Health". We further extracted 112,090 questions (including their titles and contents) from one of the sub-categories "Alternative Medicine". We then matched the preferred DS names in "iDISK", the first Integrated DIetary Supplements Knowledge base where DS related information is represented in a comprehensive and standardized form [19], with the DS ingredient name in the questions. After two assessors (YW, RR) had manually reviewed the matched preferred names, we cleaned up the DS ingredient names list based on the following rules: 1) only including ingredients with more than 5 matched questions; 2) excluding commonly consumed everyday food/drink items, e.g., fruits, vegetables, wine, caffeine, and water; 3) excluding body parts, e.g., adrenal cortex, brain, and stomach; and 4) excluding recreational drugs e.g., marijuana, poppy seed. Only the questions that exactly matched the DS ingredient names on this list were kept.

These questions were further pre-processed by subject matter experts (TN, JV) to be used for topic modeling. We removed all ingredient mentions within the questions to understand the information needs non-specific to certain DS. Each question was then lower-cased and tokenized. Special characters, hyperlinks, and common stop-words (e.g., 'I', 'you', etc.) were removed, and each word was normalized using the normalized string generator (Norm) from the SPECIALIST NLP tool [20]. We only considered words that had at least 3 characters, since any word shorter than that was usually not meaningful. We also removed words that occurred fewer than five times, or more than 85% of the time, as they might not contribute much to the question.

### Identifying Topics for DS Questions

In our preliminary investigation of different topic modeling strategies, we found that Correlation Explanation (CoRex) [21] discovered the most coherent topics compared to Latent Dirichlet Allocation (LDA). In contrast to LDA, which defines a generative model for inferring topics, CoRex discovers topics by maximizing the mutual information between words and topics. A subjective assessment of topic quality was performed by two assessors/co-authors and subject matter experts (YW, RR). A topic was considered "coherent" by the experts if assessors found a clear semantic criterion that unites the words under a particular topic. In total, we evaluated several results corresponding to various CoRex models on different numbers of topics (n = 100, 150, 200, 250). Comparing topic modeling results from 100 to 250 topics, we found the model with 200 topics yields the most coherent topic categories.

The selected model was further analyzed and assigned topic names after mutual agreement between two assessors (YW, RR). The "topics" with similar themes were then merged into "categories" (e.g., gastrointestinal disorders, psychiatric disorders) that were further condensed into higher level "groups" (e.g., "uses and symptoms"). For the group, "uses and symptoms", we utilized System Organ Classification (SOC) created by the Medical Dictionary for Regulatory Activities (MedDRA), a medical terminology used to classify adverse event information associated with the use of biopharmaceuticals and other medical products [22].

### Topic Evaluation

To evaluate the accuracy of the topic modeling, we selected 15 topics and extracted their corresponding 10 questions with highest ranked probabilities. Manual review (RR, YW) was conducted to determine if the extracted questions correctly aligned with topics generated by the above topic modeling methods. The measure of correctness was reported as percentage accuracy. We also extracted the DS ingredient names corresponding to each topic in order to explore the distribution of ingredients names across various topics. We also reported the DS ingredients associated with most questions for selected topics.

## Results

### Question Data and Topic Analysis

The final list consisted of 438 unique DS terms in total associated with 16,095 unique matching questions. After data pre-processing, our corpus contained a total of 213,790 tokens, which made up of 7,164 unique words.

From the 200 topics generated by CoRex modeling method, the domain experts (RR, YW) identified topics with similar themes and classified them into 38 unique categories by (Table 1). The 38 unique categories were further summarized into the following 12 higher level groups: uses or adverse effects, product-related, healthy lifestyle, information resources/scientific evidence, addiction, time of use qualifier, sleep disorder, interventions, adverse effect in general, health benefits, mind and body, and population qualifier. The distribution of higher-level groups and number of their associated categories is provided below (Figure 2).

After evaluating the top 10 ranked questions for selected topics, we reported accuracy as number and percentage of questions that correctly aligns with the generated topic. Table 1 lists examples of selected topic groups, their associated categories along with the top 15 most probable words and common ingredients mentions.

*Table 1. Selected topic categories with associated features and accuracy of topic modeling*

| Topic Groups | Topic categories (Topic Index) | Representative key words | Question example | Number of correctly matched (accuracy) | Representative supplements |
|---|---|---|---|---|---|
| Uses & adverse effects | Gastrointestinal disorders (65) | constipation, laxative, fiber, enema, suppository, constipate, dulcolax, docusate, insoluble, poop, saline, fleet, along with, bum, fecal | Want to take magnesium for health, but it irritates my IBS? Does anyone have any advice for taking Mg? | 10 (100%) | Magnesium, Senna, Castor, Probiotics, Glycerin |
| | Musculoskeletal disorders (93) | arthritis, bracelet, rheumatoid, knee, magnetic, anabolic, steroid, juvenile, sheet, gout, rheumatism, mattress, wonderfully, bangle, supplemental | Is hyaluronic acid supplement safe for children with juvenile rheumatoid arthritis? | 10 (100%) | Copper, Fish Oil, Vitamin D |
| | Psychiatric Disorders (11) | john, anxiety, depression, wort, htp, antidepressant, calm, social, root, 5htp, sam, tryptophan, zoloft, ssri, withdrawal | Is there alternative solution for anxiousness. I'd like to know is if anybody has any feedback on Buspar, Kava root, Passion flower & St. John's wort? | 10 (100%) | Valerian, St. John's Wort, SAM-e |
| | Respiratory, thoracic/mediastinal disorders (1) | remedy, throat, sore, home, lemon, salt, hot, cure, natural, gargle, warm, epsom, voice, soothe, homeopathic | How can I soothe my sore throat? I have been drinking warm tea with honey & gargle some salt water but it doesn't work. | 9 (90%) | Honey, Ginger, Vitamin C, Garlic |
| | Skin & subcutaneous tissue disorders (23) | skin, face, acne, wash, foot, hand, red, spot, itchy, facial, cream, wrinkle, mask, oily, swell | What are some natural ways to get rid of pimples (home remedies)? I have heard about Aloe | 10 (100%) | Apple cider vinegar, Fish oil, Honey, Tea tree oil |
| | Cardio-vascular/blood & lymphatic system disorders (2) | blood, level, low, pressure, high, normal, cholesterol, , cause, heart, disease, diagnose, raise, thin, flow | Does anyone know about an alternative medicine way of treating high blood pressure? I have heard of l-arginine to be effective. | 9 (90%) | Iron, Fish Oil, Sodium |
| | Endocrine disorders (68) | thyroid, hypothyroidism, synthroid, levothyroxine, hypothyroid, gut, patient, syndrome, leaky, radioactive, testimony, underactive, iodide, success, advisable | Are there natural alternatives after having radioactive iodine treatment for Graves' disease? | 10 (100%) | Iodine, Kelp, Vitamin D |
| | Infections & infestations (31) | infection, yeast, bladder, antibiotic, treat, kill, bite, mannose, parasite, mosquito, coat, douche, poultice, intestinal, frequent | Home remedies for yeast infections? I have heard where diluted apple cider vinegar in a bath tub could provide temporary relief | 10 (100%) | Cranberry, Garlic, Apple Cider Vinegar, |
| | Pregnancy, puerperium/perinatal conditions (40) | control, birth, period, pregnancy, pregnant, miscarriage, irregular, headache, hormone, tension, 1st, abortion, fertility, defect, endometriosis | Is it safe to drink penny royal tea along with parsley tea to induce a miscarriage? | 10 (100%) | Vitamin C, Iron, Dong Quai |
| | Dental & gingival conditions (124) | pot, tooth, eliminate, desperate, prune, brush, resistant, walnut, wisdom, piss, neti, dentist, dependency, swish, tonsillitis | I am getting my wisdom teeth out & I would like to know if clove oil will help in pain in my gums? | 10 (100%) | Apple Cider Vinegar, Silver, Garlic, Clove |
| Addiction | Smokables (21) | smoke, wee, marihuana, cigarette, legal, pipe, bowl, roll, tobacco, illegal, bud, blunt, smoker, kush, hash | What is the difference between Salvia-A & Salvia divinorum? I smoked salvia divinorum | 10 (100%) | Damiana, Catnip, Mullein, Salvia divinorum, Kratom, Clove |
| Product-related | Dose/dose form/preparation (43) | mcg, 1000, complex, omega, 5000, 2000, 2500, multiple, adult, 2000iu, standardize, milligram, cla, strength, gnc | Is the dose of vitamin A 2500 IU? Why can I not find that in a supplement? | 10 (100%) | Biotin, Vitamin D, Vitamin C, Fish Oil, Vitamin B12, Folic Acid, Vitamin A |
| Sleep disorder | Sleeping (50) | fall, asleep, 2am, cry, couch, 3am, deaf, sleeper, category, pulmonary, carpet, regardless, proportion, haven't, recreational | Need more info about Melatonin? I have trouble falling asleep | 8 (80%) | Melatonin, Valerian |
| Frequency/Time | Frequency/Time reference (9) | even, month, since, may, though, well, always, never, pretty, end, fine, mean, without, course, quite | Chromium supplements… do they work? I have been taking them for a few weeks now | 7 (70%) | Iron, Vitamin D, Cranberry |
| Health life style | Weight control (19) | weight, lose, gain, loose, berry, cardio, primarily, weight, workout, lift, shed, cambogia, underweight, usda | Would it be safe to take a weight loss pill with the supplements? | 9 (90%) | Acai, Apple Cider Vinegar, Honey, Garcinia, |

We found the percent accuracy for most of the selected topics is between 90%-100%, except sleep (80%) and frequency/time categories (70%). "Use and adverse effects" is the most dominant topic group, which accounted for 50 topics out of 200. Under this topic there were 15 categories classified based on MedDRA SOC (Figure 3).

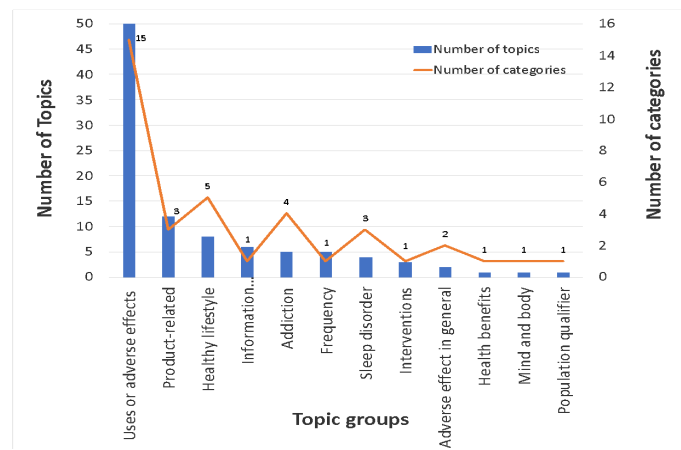

*Figure 2-Distribution of topic groups and their associated number of categories.*

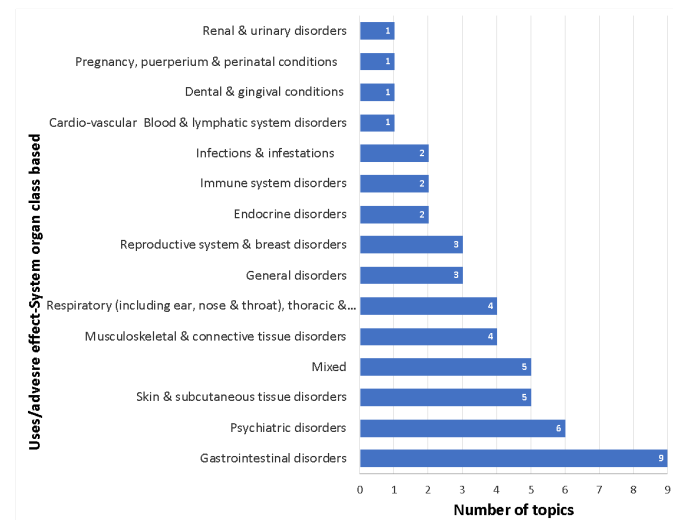

*Figure 3- Distribution of uses/adverse effects related categories based on system organ class (SOC)*

**Dietary Supplements Associated with Most Questions**

We also extracted the DS ingredient names associated with most questions corresponding to a particular topic in order to explore the distribution of most commonly discussed ingredients. Only DS ingredients associated with ≥10% of questions under a specific topic were reported (Table 2).

## Discussion

In this study, we employed CorEx topic modeling over user-generated questions coming from the Yahoo! Answers data in order to better understand the information needs of consumers. We also discovered interesting information on the distribution of DS ingredients across topics of special interest to consumers. This research effort further validates the feasibility of topic modeling to extract important information hidden in large corpus of social media data.

Applying CorEx topic modeling methods, we were able to accurately identify 12 topic groups. The top three groups with the most number of respective assigned categories and topics, which can be regarded as the information most sought by consumers, are: "use and adverse effects", "product-related", and "healthy life style" (Figure 2). Extracted information pertaining to any symptom or sign could either be an indication or an adverse event of a DS, (e.g., diarrhea, abdominal pain, palpitations, headaches). Thus uses and adverse effects were combined as one group, "use and adverse effects". We found a higher number of topics and the associated number of questions concerning: gastrointestinal system (specifically diarrhea and constipation); psychiatric (mainly anxiety and depression); and skin and subcutaneous tissues (primarily acne and UV protection). We also had a "mixed group", having keywords corresponding to more than one system. For "product-related groups", we merged categories like dose, dose from, preparation because of their co-occurrence under one topic (e.g., Topic #43). Under the "healthy life style" group, the topics were mostly around eating healthy and weight control/exercise.

*Table 2. Total number of questions matched for each topic. The representative ingredient is the one that matched the most questions. The percentage of the questions that mentioned the representative ingredient is shown in the parenthesis. The text in bold represents the ingredient with high percentage of associated questions.*

| Topic categories Topic Index) | Number of questions matched | Representative Ingredient | Questions containing Representative Ingredient |
|---|---|---|---|
| Gastrointestinal disorders (65) | 145 | Magnesium | 24 (16.55%) |
| Musculoskeletal disorders (93) | 45 | Copper | 11 (24.44%) |
| Psychiatric disorders (11) | 256 | Valerian | 45 (17.58%) |
| Respiratory (including ear, nose & throat), thoracic & mediastinal disorders (1) | **476** | **Honey** | **226 (47.48%)** |
| Skin & subcutaneous tissue disorders (23) | 84 | Apple Cider Vinegar | 17 (20.24%) |
| Cardio-vascular/blood & lymphatic system disorders (2) | 264 | Iron | 38 (14.39%) |
| Endocrine disorders (68) | 48 | Iodine | 11 (22.92%) |
| Infections & infestations (31) | 160 | Cranberry | 37 (23.12%) |
| Smokables (21) | 134 | Damiana | 11 (8.21%) |
| Dose/dose form/preparation (43) | 98 | Biotin | 35 (35.71%) |
| Sleeping (50) | **45** | **Melatonin** | **29 (64.44%)** |
| Weight control (19) | 171 | Acai | 26 (15.20%) |

We found high accuracy in identifying questions that correctly align with the topic categories/groups (Table 1). We found few low matching accuracy topics also having questions related to other topics, e.g., sleeping disorders topic with questions related to recreational drug, anxiety/depression. We also reported actual DS ingredient names associated with most questions for a particular topic (Table 2). We found a substantially higher percentage of questions for the ingredients "Honey" under respiratory disorders and "Melatonin" under sleep disorders. This information provides essential knowledge on the use of DS for various specific reasons and needs further exploration.

Several limitations were noticed. We just analyzed questions belonging to alternative medicine sub-category under "health" section and might have missed dietary supplement occurrences under other sub-categories, e.g., mental health conditions, general health care. We only used preferred DS ingredient names and not their synonyms (e.g., scientific names, common names) to extract the corresponding questions. Also, there are inherent limitations to topic modeling e.g., topics were generated based on the statistical word distribution within the questions and thus topics with incoherent topic keywords were also generated.

## Conclusions

This research provides essential insights on extracting and understanding the information needs of consumers around dietary supplements using CorEx-based topic modeling that could identify the relevant topics embedded in a large corpus of Yahoo! Answers data with high accuracy. The knowledge gained here could be used to generate a more comprehensive repository of resource for consumers around dietary supplements usage. Thus, this study is an important contribution in further accentuating the potential benefits of using social media data in the clinical research.

## Acknowledgements

This research was supported by the U.S. National Institutes for Health (R01AT009457) (Zhang).

## References

[1] E. D. Kantor, C. D. Rehm, M. Du, E. White, and E. L. Giovannucci, "Trends in Dietary Supplement Use Among US Adults From 1999-2012," (in eng), *JAMA,* vol. 316, no. 14, pp. 1464-1474, Oct 2016, doi: 10.1001/jama.2016.14403.

[2] J. J. Gahche, R. L. Bailey, N. Potischman, and J. T. Dwyer, "Dietary Supplement Use Was Very High among Older Adults in the United States in 2011-2014," (in eng), *J Nutr,* Aug 2017, doi: 10.3945/jn.117.255984.

[3] D. J. Amante, T. P. Hogan, S. L. Pagoto, T. M. English, and K. L. Lapane, "Access to care and use of the Internet to search for health information: results from the US National Health Interview Survey," *Journal of medical Internet research,* vol. 17, no. 4, 2015.

[4] S. Fox. "Health topics: 80% of internet users look for health information online. Washington, DC: Pew Internet & American Life Project." http://www.pewinternet.org/2005/05/17/health-information-online/ (accessed.

[5] P. Institute of Medicine . Committee on the Use of Complementary and Alternative Medicine by the American, *Complementary and alternative medicine in the United States*. Washington, D.C. National Academies Press, 2005.

[6] "FDA 101: Dietary Supplements." https://www.fda.gov/ForConsumers/ConsumerUpdates/ucm050803.htm (accessed 2017).

[7] "National Institute Of Health (NIH), Office of Dietary Supplements (ODS)." https://ods.od.nih.gov (accessed 2013).

[8] "Dietary Supplement Label Database (DSLD)." https://www.dsld.nlm.nih.gov/dsld/index.jsp (accessed.

[9] "Natural Medicine (NM)." https://naturalmedicines.therapeuticresearch.com (accessed.

[10] D. M. Blei, "Probabilistic topic models," *Communications of the ACM,* vol. 55, no. 4, pp. 77-84, 2012.

[11] L. Wang, J. Lakin, C. Riley, Z. Korach, L. N. Frain, and L. Zhou, "Disease Trajectories and End-of-Life Care for Dementias: Latent Topic Modeling and Trend Analysis Using Clinical Notes," AMIA 2018 Annual Symposium, 2018.

[12] A. L. Beam *et al.*, "Predictive modeling of physician-patient dynamics that influence sleep medication prescriptions and clinical decision-making," *Scientific reports,* vol. 7, p. 42282, 2017.

[13] Y. Wang, D. R. Gunashekar, T. J. Adam, and R. Zhang, "Mining Adverse Events of Dietary Supplements from Product Labels by Topic Modeling," *Studies in health technology and informatics,* vol. 245, p. 614, 2017.

[14] M. J. Paul and M. Dredze, "Discovering health topics in social media using topic models," *PloS one,* vol. 9, no. 8, p. e103408, 2014.

[15] K. W. Prier, M. S. Smith, C. Giraud-Carrier, and C. L. Hanson, "Identifying health-related topics on twitter," in *International conference on social computing, behavioral-cultural modeling, and prediction*, 2011: Springer, pp. 18-25.

[16] Z. He, Z. Chen, S. Oh, J. Hou, and J. Bian, "Enriching consumer health vocabulary through mining a social Q&A site: A similarity-based approach," *Journal of biomedical informatics,* vol. 69, pp. 75-85, 2017.

[17] M. S. Park, Z. He, Z. Chen, S. Oh, and J. Bian, "Consumers' use of UMLS concepts on social media: diabetes-related textual data analysis in blog and social Q&A sites," *JMIR medical informatics,* vol. 4, no. 4, 2016.

[18] Zhang Z, Yu L, Kou Y, Wu D, He Z, "Understanding Patient Information Needs about their Clinical Laboratory Results," *Studies in Health Technology and Informatic*. In press.

[19] X. He *et al.*, "Prototyping an Interactive Visualization of Dietary Supplement Knowledge Graph," in *2018 IEEE International Conference on Bioinformatics and Biomedicine (BIBM)*, 2018: IEEE, pp. 1649-1652.

[20] "Lexical tools." https://lexsrv3.nlm.nih.gov/LexSysGroup/Projects/lvg/2013/docs/userDoc/tools/norm.html (accessed 2011).

[21] G. Ver Steeg and A. Galstyan, "Discovering structure in high-dimensional data through correlation explanation," in *Advances in Neural Information Processing Systems*, 2014, pp. 577-585.

[22] "Medical Dictionary of Regulatory Activities (MedDRA)." https://www.meddra.org/about-meddra/quality-controls (accessed 2018).

**Address for correspondence**
Rui Zhang, PhD, Email: zhan1386@umn.edu